\documentclass[12pt]{article}
\usepackage{graphicx}
\usepackage{color}
\usepackage{cite}
\usepackage{amsmath}

\def\upstrut{\vrule height 2.5ex depth 0.0ex width 0pt}
\def\medstrut{\vrule height 2.5ex depth 1.0ex width 0pt}

\def\ts{\textstyle}
\def\css{\hbox{$c\kern0.03em(\kern-0.1em ss\kern-0.1em)$}}
\def\bss{\hbox{$b\kern0.03em(\kern-0.1em ss\kern-0.1em)$}}
\def\qss{\hbox{$q\kern0.03em(\kern-0.1em ss\kern-0.1em)$}}
\def \barM{\overline{M}}
\def \beq{\begin{equation}}
\def \eeq{\end{equation}}
\def\eqref#1{(\ref{#1})}
\def\bea{\begin{eqnarray}}
\def\eea{\end{eqnarray}}
\def \bra#1{\langle{#1}|}
\def \ket#1{|{#1}\rangle}
\def\jpsi{J\kern-0.15em/\kern-0.15em\psi\kern0.15em}
\def\SP{\hbox{$S$-$P$}\ \strut}

\def\Sd{S_{\kern-0.1em ss}}
\def\SSd{{\bold S}_{\kern-0.1em ss}}
\def\Sdi{S^i_{\kern-0.1em ss}}
\def\SQj{S_Q^j}
\def\third{\hbox{$\frac{1}{3}$}}
\def\nl{\hfill\break}
\def\lto{\longrightarrow}

\def\URLtilde{\lower0.2em\hbox{$\tilde{\phantom{a}}$}}
\def\mycomm#1{\hfill\break\strut\kern-3em{\color{red}\tt ====> #1
\color{black}}\hfill\break}

\newcount\timecount
\newcount\hours \newcount\minutes  \newcount\temp \newcount\pmhours
\hours = \time
\divide\hours by 60
\temp = \hours
\multiply\temp by 60
\minutes = \time
\advance\minutes by -\temp
\def\hour{\the\hours}
\def\minute{\ifnum\minutes<10 0\the\minutes
\else\the\minutes\fi}
\def\clock{
\ifnum\hours=0 12:\minute\ AM
\else\ifnum\hours<12 \hour:\minute\ AM
\else\ifnum\hours=12 12:\minute\ PM
\else\ifnum\hours>12
\pmhours=\hours
\advance\pmhours by -12
\the\pmhours:\minute\ PM
\fi
\fi
\fi
\fi
}

\def\monthname{\relax\ifcase\month 0/\or January\or February\or
March\or April\or May\or June\or July\or August\or September\or
October\or November\or December\else\number\month/\fi}
\def\today{\monthname~\number\day, \number\year}

\def\bold#1{\boldsymbol{#1}}

\def\draft{\color{red}
$\bold{\strut\kern-3em
\hbox{\tt \Large DRAFT, NOT TO BE DISTRIBUTED:  \clock, \today.}
}$\par\noindent\color{black}}
\begin{document}
\setcounter{footnote}{1}
\rightline{EFI 17-8}
\rightline{TAUP 3016/17}
\rightline{arXiv:1703.07774}
\vskip1.5cm
\centerline{\large \bf \boldmath VERY NARROW EXCITED $\Omega_c$ BARYONS
\unboldmath}
\bigskip

\centerline{Marek Karliner$^a$\footnote{{\tt marek@proton.tau.ac.il}},
and Jonathan L. Rosner$^b$\footnote{{\tt rosner@hep.uchicago.edu}}}
\medskip

\centerline{$^a$ {\it School of Physics and Astronomy}}
\centerline{\it Raymond and Beverly Sackler Faculty of Exact Sciences}
\centerline{\it Tel Aviv University, Tel Aviv 69978, Israel}
\medskip

\centerline{$^b$ {\it Enrico Fermi Institute and Department of Physics}}
\centerline{\it University of Chicago, 5620 S. Ellis Avenue, Chicago, IL
60637, USA}
\bigskip
\strut

\begin{center}
ABSTRACT
\end{center}
\begin{quote}
Recently LHCb reported the discovery of five extremely narrow excited
$\Omega_c$ baryons decaying into $\Xi_c^+ K^-$.  We interpret these baryons as
bound states of a $c$-quark and a $P$-wave $ss$-diquark. For such a system
there are exactly five possible combinations of spin and orbital angular
momentum.  The narrowness of the states could be a signal that it is hard to
pull apart the two $s$-quarks in a diquark.  We predict two of spin 1/2, two of
spin 3/2, and one of spin 5/2, all with negative parity.
Of the five states two can decay in $S$-wave and three can decay in $D$-wave.
Some of the $D$-wave states might be narrower than the $S$-wave states.  We
discuss relations among the five masses expected in the quark model and the
likely spin assignments and compare with the data.
A similar pattern is expected for negative-parity excited $\Omega_b$ states.
An alternative interpretation is noted in which the heaviest two states are
$2S$ excitations with $J^P = 1/2^+$ and $3/2^+$, while the lightest three are
those with $J^P = 3/2^-,3/2^-,5/2^-$ expected to decay via $D$-waves.  In
this case we expect $J^P = 1/2^-$ $\Omega_c$ states around 2904 and 2978 MeV.
\end{quote}
\smallskip

\leftline{PACS codes: 12.39.Jh, 13.20.Jf, 13.25.Jx, 14.40.Rt}

\vfill\eject

\section{Introduction \label{sec:intro}}

Very recently LHCb reported the discovery of five extremely narrow excited
$\Omega_c$ baryons decaying into $\Xi_c^+ K^-$ \cite{LHCbOmegac},
with masses and widths shown in Table \ref{tab:omc}.  We quote also our
favored spin-parity assignment for these states, which we shall choose
among the 5! = 120 possible permutations if all five states are $P$-wave
excitations of the $ss$ diquark with respect to the charmed quark.
Some more recent calculations \cite{Padmanath:2017,Wang:2017vnc,Wang:2017zjw,%
Chen:2017gnu,Aliev:2017led}
that appeared after the first version of this paper reach the same
conclusion.  In parentheses we note an alternative assignment if the two
heaviest states are $2S$ excitations.

\begin{table}
\caption{Masses and widths of $\Omega_c = css$ candidates reported by the
LHCb Collaboration \cite{LHCbOmegac}.  The proposed values of spin-parity $J^P$
are ours.  An alternative set of assignments is shown in parentheses.
\label{tab:omc}}
\begin{center}
\begin{tabular}{c c c c} \hline \hline
State & Mass (MeV)$^a$ & Width (MeV) & Proposed $J^P$ \\ \hline
$\Omega_c(3000)^0$ & $3000.4 \pm 0.2 \pm 0.1$ & $4.5\pm0.6\pm0.3$ & $1/2^-$
($3/2^-$) \\
$\Omega_c(3050)^0$ & $3050.2 \pm 0.1 \pm 0.1$ & $0.8\pm0.2\pm0.1$ & $1/2^-$
($3/2^-$) \\
&& $< 1.2$ MeV, 95\% CL & \\
$\Omega_c(3066)^0$ & $3065.6 \pm 0.1 \pm 0.3$ & $3.5\pm0.4\pm0.2$ & $3/2^-$
($5/2^-$) \\
$\Omega_c(3090)^0$ & $3090.2 \pm 0.3 \pm 0.5$ & $8.7\pm1.0\pm0.8$ & $3/2^-$
($1/2^+$) \\
$\Omega_c(3119)^0$ & $3119.1 \pm 0.3 \pm 0.9$ & $1.1\pm0.8\pm0.4$ & $5/2^-$
($3/2^+$) \\
&& $< 2.6$ MeV, 95\% CL & \\
\hline \hline
\end{tabular}
\end{center}
\leftline{$^a$Additional common error of +0.3,--0.5 MeV from $M(\Xi_c^+)$
uncertainty.}
\end{table}

This discovery raises some immediate questions, which we address in detail: 
\begin{itemize}
\item[(a)] Why five states?  Are there more in this $css$ system?
\item[(b)] Why are they so narrow?
\item[(c)] What are their spin-parity assignments?
\item[(d)] Can one understand the mass pattern?
\item[(e)] Are there other similar states with different quark content,
in particular very narrow excited $\Omega_b$ baryons?
\end{itemize}

In Sec.\ \ref{sec:PW} we comment on $P$-wave $css$ baryons.  We then
analyze spin-dependent forces for the $css$ system in Sec.\ \ref{sec:sd},
building upon similar results \cite{Karliner:2015ema} obtained previously
for the negative-parity $\Sigma_c$ states.  We evaluate the energy cost for a
$P$-wave $css$ excitation in Sec.\ \ref{sec:SP}, carry our results over to the
$\Omega_b$ system in Sec.\ \ref{sec:omb}, discuss alternative interpretations
of the spectrum in Sec.\ \ref{sec:alt}, and conclude in Sec.\ \ref{sec:concl}.
Details of calculating the spin-dependent mass shifts are presented in 
Appendix A, with a linearized approximation in Appendix B.

\section{\boldmath $P$-wave \css\ system \unboldmath \label{sec:PW}}

Consider the $(ss)$ in \css\ to be an $S$-wave color ${\bf\bar 3}_c$ 
diquark.  Then it must have spin $\Sd = 1$.  This spin can be combined
with the spin 1/2 of the charm quark $c$ to a total spin $S=1/2$ or 3/2.
Consider states with relative orbital angular momentum $L=1$ between the
spin-1 diquark and the charm quark.  Combining $L=1$ with $S = 1/2$ we get
states with total spin $J=1/2,3/2$, while combining $L=1$ with $S=3/2$ we get
states with $J=1/2,3/2,5/2$.  All five states have negative parity $P$.  Those
with $J^P=1/2^-$ decay to $\Xi_c^+ K^-$ in an $S$-wave, while those with $J^P =
3/2^-,5/2^-$ decay to $\Xi_c^+ K^-$ in a $D$-wave.

Two states with the same $J^P$ could interfere with one another, but the line
shapes of the resonances do not reflect significant interference effects.  To
test possible interference effects, LHCb added an extra phase between any pair
of close peaks under the assumption they have the same quantum numbers.
The effect of the interference turned to be negligible \cite{MPPC}.

The narrowness of the states could be a signal that it is hard to pull apart
the two $s$ quarks in a diquark.  One $s$ quark has to go into the $K^-$ and
the other into the $\Xi_c^+$.  It is also possible that the three states which
have to decay by a $D$-wave are narrower than the other two.  We shall find
that our preferred $J^P$ assignments only partially conform to this
expectation, while an alternative assignment is consistent with it.

If indeed the narrowness of these $\Omega_c$ states is due to the difficulty
of pulling apart the two quarks in an $(ss)$ diquark, then perhaps this
can also explain the narrowness of some excited ordinary $\Xi$ baryons. The
analogy is as follows.

\beq
\strut\kern-6em
\begin{array}{ccccc}
&\Omega_c      & \lto & \Xi_c^+ &  K^-  \\
&c\hbox{-}(ss) &      & (csu)  & (s \bar u)  \\
\hbox{Replacing $c$ by $u$:\qquad} &&&& \\
&u\hbox{-}(ss) &      & (usu)  & (s \bar u)  \\
&\Xi^0      & \lto & \Sigma^+ &  K^-  
\end{array}
\eeq

There are several excited $\Xi$ baryons whose decay channels include
$\Sigma \bar K$ and $\Lambda \bar K$ and which have quite narrow widths,
even though some of them have large phase space available for the decay
\cite{PDG}:\nl\nl\noindent
$\Xi(1690), \Gamma < 30$ MeV ($J^P$ unknown),\nl
$\Xi(1820), \Gamma = 24^{{+}16}_{{-}10}$ MeV ($J^P=3/2^-$),\nl
$\Xi(1950), \Gamma = 60 \pm 20$ MeV ($J^P$ unknown),\nl
$\Xi(2030), \Gamma = 20^{{+}15}_{{-}5}$ MeV ($J^P$ unknown).
\nl

The analogy is only partially correct, because in the $csq$ system there is no
analogue of the relatively light $\Lambda$ which is 77 MeV lighter than
$\Sigma^0$.  The $\Lambda$ contains a $ud$ $I=0$ spin-0 diquark which
is significantly lighter than the $I=1$ $ud$ spin-1 diquark in $\Sigma^0$.
Under $c \leftrightarrow u$,
$\Lambda (sud)$ is replaced by $(scd) = \Xi_c^0$. In the latter 
the corresponding spin-0 $(cd)$ diquark has no reason to be light.
Perhaps this explains why the $\Omega_c$ states are significantly more
narrow than the $\Xi$ states.

In this context note that the {\em only} way for \hbox{$c$$(ss)$} states below
a certain mass to decay hadronically is to rip apart the two $s$ quarks in
an $ss$ diquark.\footnote{Isospin-violating decay into $\Omega_c \pi^0$ is
possible but highly suppressed, as discussed in Sec. VI.}
The alternative is kinematically forbidden:  if the two $s$ quarks
remain together, than the decay is $\css\lto \qss\,\,(c\bar q)$,
i.e., the final state is \ $\Xi D^{(*)}$. The lightest among these is
\ $\Xi^0 D^0$ at 3180 MeV,
which is 61 MeV above the heaviest of the narrow states, $\Omega_c(3119)$.

\section{Spin-dependence of masses \label{sec:sd}}

We recapitulate the discussion in Ref.\ \cite{Karliner:2015ema},
replacing the spin-1, isospin-1 $(uu,ud,dd)$ diquark with a spin-1,
isospin-0 doubly strange diquark $(ss)$.  We adopt the notation of Ref.\
\cite{Ebert:2011kk}, who have predictions for the masses
of these states which we shall discuss presently.  The spin-dependent potential
between a heavy quark $Q$ and the $(ss)$ spin-1 diquark is
\beq \label{eqn:vsd}
V_{SD} = a_1{\bold L} \cdot \SSd + a_2{\bold L} \cdot {\bold S_Q}
 + b [ - \SSd \cdot {\bold S_Q}
 + 3(\SSd \cdot {\bold r})({\bold S_Q} \cdot {\bold r})/r^2]
 + c \SSd \cdot {\bold S_Q}~,
\eeq
where the first two terms are spin-orbit forces, the third is a tensor force,
and the last describes hyperfine splitting.  If $a_1 = a_2$, the spin-orbit
force becomes proportional to ${\bold L}\cdot (\SSd + {\bold S_Q})=
{\bold L} \cdot {\bold S}$, where ${\bold S}$ is the total spin, so states may
be classified as $^{2S+1}P_J =$\\
$ ^2P_{1/2},~^2P_{3/2},~^4P_{1/2},~^4P_{3/2},$ and $^4P_{5/2}$.  When $a_1 \ne
a_2$, the states with the same $J$ but different $S$ mix with
one another and are eigenstates of $2 \times 2$ matrices, involving a
correction to Ref.\ \cite{Karliner:2015ema} (see also \cite{LL}).  Details
of this calculation are given in Appendix A.

\beq \label{eqn:m12}
\Delta {\cal M}_{1/2} = \left[ \begin{array}{c c} \frac13 a_2 - \frac43 a_1 &
\frac{\sqrt{2}}{3} (a_2-a_1) \\ \frac{\sqrt{2}}{3}(a_2-a_1) &
 - \frac53 a_1 - \frac56 a_2
\end{array} \right] +b \left[ \begin{array}{c c} 0 & \frac{1}{\sqrt{2}} \\
\frac{1}{\sqrt{2}}& -1 \end{array} \right] + c \left[ \begin{array}{c c} -1 &
 0 \\ 0 & \frac12  \end{array} \right]~,
\eeq
\beq \label{eqn:m32}
\Delta {\cal M}_{3/2} = \left[ \begin{array}{c c} \frac23 a_1 - \frac16 a_2 &
\frac{\sqrt{5}}{3}(a_2-a_1) \\ \frac{\sqrt{5}}{3}(a_2-a_1) &
 - \frac23 a_1 - \frac13 a_2
\end{array} \right] +b \left[ \begin{array}{c c} 0 & -\sqrt{5}/10 \\
 -\sqrt{5}/10 & \frac45 \end{array} \right] + c \left[ \begin{array}{c c} -1 &
 0 \\ 0 & \frac12 \end{array} \right]~,
\eeq
\beq \label{eqn:m52}
\Delta {\cal M}_{5/2} = a_1 + \frac12 a_2 - \frac15 b + \frac12 c~.
\eeq
The spin-weighted sum of these mass shifts is zero:
\beq \label{eqn:wtsum}
\sum_{J} (2J+1) \Delta {\cal M}_J = 0
\eeq
implying one linear relation among the mass shifts.  For any given assignment
of the five states to two values of $J^P = 1/2^-$, two of $J^P = 3/2^-$, and
one of $J^P = 5/2$, there should, in principle, exist exactly one solution
for the four parameters $a_1,~a_2,~b$, and $c$. 
In practice, as we discuss below,
only one solution in which all states are $P$-waves gives reasonable values of
these parameters, and it is the one shown in Table \ref{tab:omc}.

Although $m_c$ is not much larger than the $(ss)$ diquark mass (which we
shall evaluate presently), it will be helpful to quote a linearized version
of the mixing using lowest-order perturbation theory in the inverse of $m_c$ 
\cite{Karliner:2015ema}.  For this one couples the $(ss)$ diquark spin
$S_{(ss)} = 1$ and the orbital angular momentum $L = 1$ to a light-quark total
angular momentum $j=0,1,2$.  The states with definite $J,j$ can be expressed in
terms of those with definite $J,S$ via Clebsch-Gordan coefficients.
Details are given in Appendix B.
Expanding in definite-\kern-0.1em$j$ eigenfunctions of the ${\bold L} \cdot \SSd$
term, the result is
\bea
\Delta M(J=\frac12,j=0) &=& -2a_1~, \label{eqn:10}\\
\Delta M(J=\frac12,j=1) &=& -a_1 -\frac12 a_2 -b -\frac12 c~,
\label{eqn:11} \\
\Delta M(J=\frac32,j=1) &=& -a_1 +\frac14 a_2 + \frac12 b + \frac14 c~,
\label{eqn:31} \\
\Delta M(J=\frac32,j=2) &=& a_1 -\frac34 a_2 +\frac{3}{10} b - \frac34 c~,
\label{eqn:32} \\
\Delta M(J=\frac52,j=2) &=& a_1 +\frac12 a_2 - \frac15 b + \frac12 c~.
\label{eqn:52}
\eea
This expresses five mass shifts in terms of four parameters.  One linear
relation among them is the vanishing of their spin-weighted sum, as before.
But here, $a_2$ and $c$ always occur in the combination $a_2 + c$, so that
the five mass shifts are expressed in terms of the three free parameters $a_1$,
$a_2+c$, and $b$.  Hence the masses satisfy one additional linear relation,
which is convenient to write as two separate ones:
\beq \label{eqn:lin1}
2 M(1/2,1) + 4 M(3/2,1) = 3 M(1/2,0)~,
\eeq
\beq \label{eqn:lin2}
4 M(3/2,2) + 6 M(5/2,2) = -5 M(1/2,0)~.
\eeq
Here the first number refers to $J$ and the second to $j$.  These two relations
imply Eq.\ (\ref{eqn:wtsum}).  They are not well satisfied by our favored
assignment, implying a shortcoming of the $1/m_c$ expansion for such a heavy
``light diquark'' $(ss)$.  We shall label states by their total $J$ and their
heavy-quark-limit $j$, even when mixed [Eqs.~\eqref{eqn:m12}--\eqref{eqn:m32}].

An initial effort to assign $J^P$ values to the five states made use of
the linearized equations (\ref{eqn:10}-\ref{eqn:52}).  With $a_1$ and
$a_2$ extrapolated from Ref.\ \cite{Karliner:2015ema}, it was shown that
$M(1/2,0) < M(1/2,1) < M(3/2,2) < M(5/2,2)$ for all reasonable values of
the tensor-force parameter $b$, while $M(3/2,1)$ could lie below all, three,
or two of the above four.  Although the pattern should be somewhat different
for the $css$ system, this greatly simplified the search for a reasonable
permutation of $J^P$ assignments.  The criteria for ``reasonable'' included
the following:

\begin{itemize}

\item[(i)] The hyperfine splitting parameter $c$ should be small, as it
depends on a $P$-wave wave function near the origin.

\item[(ii)] The parameter $a_2$ should be close to that estimated in Ref.\
\cite{Karliner:2015ema} from the $\Lambda_c$ system, $a_2 = 23.9$ MeV, as
it refers to the matrix element of a term ${\bold L} \cdot {\bold S_Q}$.

\item[(iii)] The parameter $a_1$ should be positive but smaller than the
value of 55.1 MeV estimated in Ref.\ \cite{Karliner:2015ema} as the
coefficient of the ${\bold L} \cdot {\bold S_{(uu)}}$ term.  Naive scaling
by the ratio of diquark masses would yield for the $\Omega_c$ system $a_1 =
(783/1095) \cdot 55.1 = 39.4$ MeV, where the $(uu)$ diquark mass was evaluated
in Ref.\ \cite{Karliner:2015ema}, and the $(ss)$ diquark mass is evaluated in
the next Section. 

\end{itemize}

\begin{figure}
\begin{center}
\includegraphics[width=0.95\textwidth]{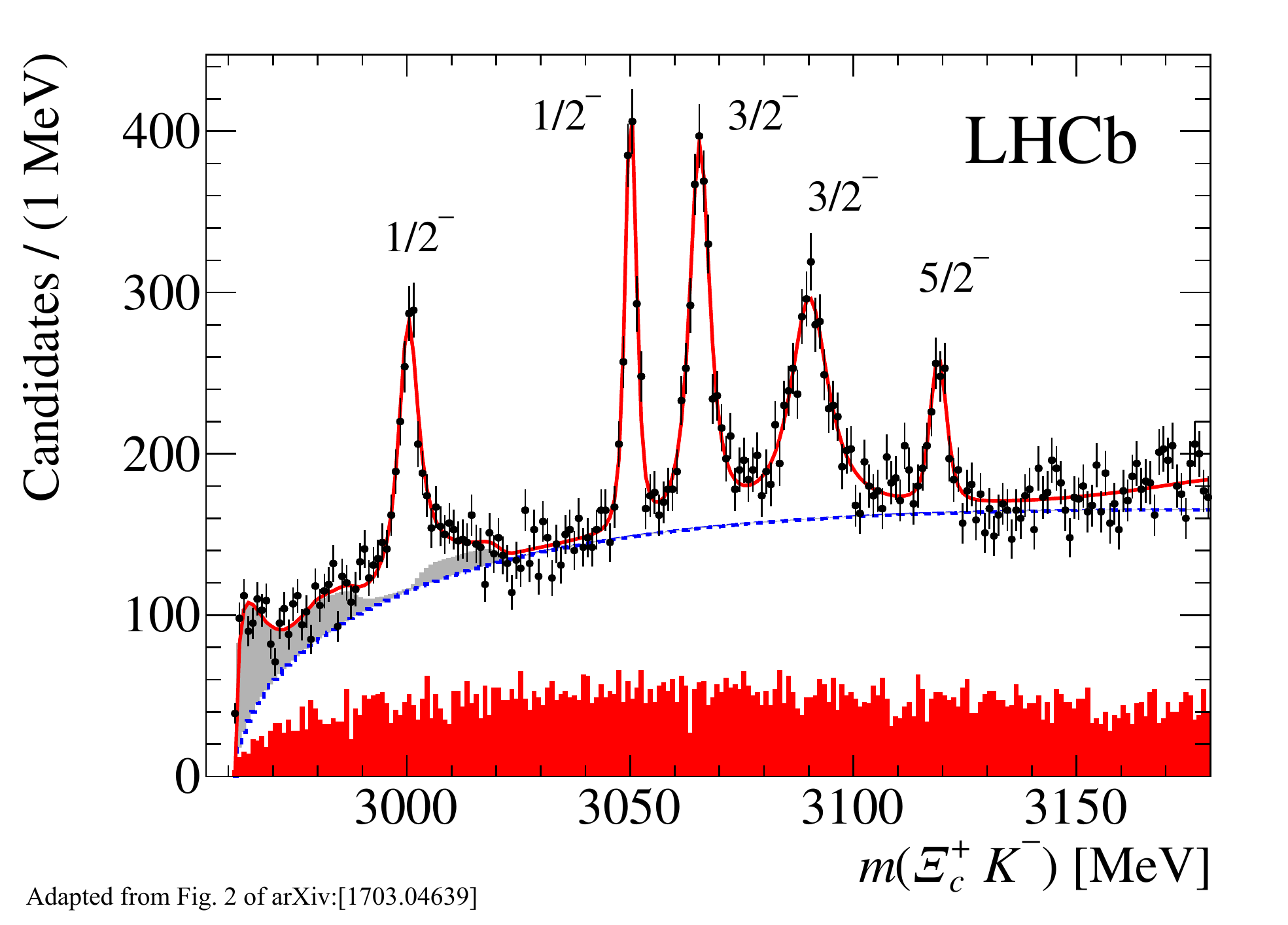}
\end{center}
\caption{\label{fig:xick} Proposed assignment of spins and parities of excited
$\Omega_c = \css$ states observed by the LHCb Collaboration if all five are
$P$-wave excitations of the $(ss)$ diquark with respect to the charmed quark.
Adapted from a zoom in on Fig.~2 of Ref.~\cite{LHCbOmegac}.}
\end{figure}

With these criteria, all $5!=120$ {\em a priori} possible assignments of
$P$-wave states were examined.  The assignment in Table \ref{tab:omc} was
favored, corresponding to the parameter choices
\beq
\boxed{
a_1 = 26.95 {\rm~MeV}~,~~
a_2 = 25.74 {\rm~MeV}~,~~
 b  = 13.52 {\rm~MeV}~,~~
 c  =  4.07 {\rm~MeV}~.
}
\eeq
This assignment of spins and parities is superposed on the LHCb
$M(\Xi_c^+ K^-)$ spectrum \cite{LHCbOmegac} in Fig.\ \ref{fig:xick}.
With the assignment in Table \ref{tab:omc}, the spin-averaged mass is
\beq
\overline{M} = (1/18) \sum_J(2J+1)M(J) = 3079.94~{\rm MeV}~.
\label{eq:Mbar}
\eeq
However, the sum rules (\ref{eqn:lin1})  and (\ref{eqn:lin2}) are poorly
obeyed, showing the shortcoming of the linear approximation for the
     \css\ system.

One other plausible assignment consists of interchanging
the states at 3050 and 3066 MeV, giving rise to a parameter set
\beq
a_1 = 21.40 {\rm~MeV}~,~~
a_2 = 40.75 {\rm~MeV}~,~~
 b  =  5.67 {\rm~MeV}~,~~
 c  =  0.45 {\rm~MeV}~.
\eeq
Here $a_1$ and $a_2$ are both farther from the expected values. 
One additional possibility involves the identification of $M(1/2,0),~M(1/2,1),
~M(3/2,1),~M(3/2,2),~M(5/2,2)$ with the respective states at 3000, 3050, 3066,
3119, and 3090 MeV.  This gives rise to a parameter set
\beq
a_1 =  21.51, {\rm~MeV}~,~~
a_2 =  -2.81 {\rm~MeV}~,~~
 b  =  38.42 {\rm~MeV}~,~~
 c  =   2.30 {\rm~MeV}~,
\eeq
with $a_2$ very far from expectations.
All the other 5! permutations lead to no solution or to
ones with negative (unacceptable) signs of $a_1$ and $a_2$.

One might have speculated that the $J=3/2$ and $J=5/2$ states, decaying to
$\Xi_c^+ K^-$ via a $D$-wave, would be narrower that those with $J=1/2$.
With our assignments, the state at 3050 MeV, assigned by us to
$J=1/2$, is seemingly the narrowest of all. But given the large statistical
error in the width of $\Omega(3119)$, LHCb cannot currently rule out the
possibility that $\Omega(3119)$ is narrower than $\Omega(3050)$ \cite{MPPC}.
No permutation of assignments which
assigns the widest states, those at 3000 and 3090 MeV, to $J=1/2$ leads to
an acceptable set of parameters.  Hence some other source of suppression of
the width of the state at 3050 MeV must be found if it really has $J=1/2$.

\section{\boldmath
Energy cost of a $P$-wave excitation of $(ss)$ diquark relative to $c$
\unboldmath
\label{sec:SP}}

The experimental \SP splitting, given the preferred spin assignments in
Table \ref{tab:omc}, is calculated from the mean \css\ $P$-wave mass,
$\overline{M} = 3079.94$ MeV, minus the spin-weighted average of the
$S$-wave masses \cite{PDG}:
\beq
(1/3)[M(\Omega_c) + 2 M(\Omega_c^*)] = (1/3)[(2695.2) + 2 (2765.9)] = 2742.33
{\rm~MeV}~,
\label{Omega_c_spin_average}
\eeq
or $\Delta E_{PS}(\Omega_c) = 337.6$ MeV.  One may ask if this is a reasonable
value.

In Ref.\ \cite{Karliner:2015ema} the corresponding splitting for $\Sigma_c$
states was estimated in Table III to be 290.7 MeV.  The reduced mass in
the $c(uu)$ system, where $(uu)$ denotes the nonstrange spin-1, isospin-1
diquark, was found to be 536.8 MeV.  The \SP splitting is expected to be a
monotonically decreasing function of reduced mass.  For a $(ss)$ diquark,
using parameters from Table I of Ref.\ \cite{Karliner:2016zzc}, one calculates
the mass of the $(ss)$ diquark to be $M_{(ss)}=2m_s^b + a/(m_s^b)^2=
2{\cdot}536.3 + 49.3{\cdot}(363.7/536.3)^2 = 1095$~MeV
and hence the reduced mass (using $m_c = 1709$ MeV) to be 667 MeV.  Using
Fig.~1 of Ref.\ \cite{Karliner:2015ema}, one would then estimate $\Delta
E_{PS}(\Omega_c) \simeq 240$ MeV, or nearly 100 MeV below the observed value.
If that were the case, at least some of the states we predict would not
correspond to the five observed by LHCb, but would lie below $\Xi_c^+ K^-$
threshold.  There is, in fact, some hint in the LHCb data just near threshold,
of some activity exceeding phase space \cite{LHCbOmegac}.\footnote{LHCb 
currently interpret the threshold enhancement as a feed-down from
$\Omega_c(3066) \to \Xi_c^\prime K \to \Xi_c \gamma K$ with the
$\gamma$ not reconstructed, but alternative interpretations, such as
additional states, are not ruled out \cite{MPPC}.}

A value of $\Delta E_{PS}(\Omega_c)$ larger than 240 MeV is estimated by
comparison with the
observed \SP splitting in the $\Xi_c$ states. The light $S$-wave, color $3^*$
diquarks $sq$ can exist in both the flavor-antisymmetric spin-0 state $[sq]$
and the flavor-symmetric spin-1 state $(sq)$.  This classification ignores
small mixing effects due to flavor-SU(3) breaking. The $S$-wave positive-parity
ground states and candidates for their $P$-wave partners are summarized in Table
\ref{tab:xic}.


\begin{table}
\caption{Lowest-lying $\Xi_c$ states classified in Ref.\ \cite{PDG} with three
stars (***).
\label{tab:xic}}
\begin{center}
\begin{tabular}{c c c c} \hline \hline
  State     &         Mass (MeV)        &  Light  &Candidate\\
            & (PDG fit or average)      & diquark &  $J^P$  \\ \hline
\medstrut
 $\Xi_c^+$  & $2467.93^{+0.28}_{-0.40}$ &  $[sq]$ & $1/2^+$ \\
\medstrut
 $\Xi_c^0$  & $2470.85^{+0.28}_{-0.40}$ &  $[sq]$ & $1/2^+$ \\
  Average   &          2469.4           &  $[sq]$ & $1/2^+$ \\
 $(\Xi')_c^+$ &    $2575.7 \pm 3.0$     &  $(sq)$ & $1/2^+$ \\
 $(\Xi')_c^0$ &    $2577.9 \pm 2.9$     &  $(sq)$ & $1/2^+$ \\
  Average   &          2576.8           &  $(sq)$ & $1/2^+$ \\
 $\Xi_c^*$  &     $2645.9 \pm 0.5$      &  $(sq)$ & $3/2^+$ \\
$\Xi_c(2790)$ &   $2789.1 \pm 3.2$      &  $[sq]$ & $1/2^-$ \\
$\Xi_c(2815)$ &   $2816.6 \pm 0.9$      &  $[sq]$ & $3/2^-$ \\
$\Xi_c(2970)^a$ &   $2970.2 \pm 2.2$      &  $(sq)$ &  $?^-$  \\
$\Xi_c(3055)^a$ &   $3055.1 \pm 1.7$      &  $(sq)$ &  $?^-$  \\
$\Xi_c(3080)^a$ &  $3076.94 \pm 0.28$     &  $(sq)$ &  $?^-$  \\ \hline \hline
\end{tabular}
\end{center}
\leftline{$^a$ The parity of these states is not yet verified
experimentally.}
\leftline{In addition PDG quotes single-\lower0.3em\hbox{*} 
\,$\Xi_c$ candidates at $2931 \pm 3 \pm
5$ and $3122.9 \pm 1.3 \pm 0.3$ MeV,}
\leftline{ which could account for the remaining $(sq)$ candidates.}
\end{table}
Only three of the five expected $c(sq)$ states are firmly established, and
we do not have spins for any of them, so we cannot use them to estimate the
\SP splitting.  However, the mass difference between the ground-state $\Xi_c
= c[sq]$ at an isospin-averaged mass of 2469.4 MeV and the spin-weighted
average of the $\Xi_c(2790)$ and $\Xi_c(2815)$ masses,
\beq
\overline{M}(c[sq],L=1) = 
   (1/3){\cdot}\left(2789.1 + 2{\cdot}2816.6\right) = 2807.4 {\rm~MeV}~,
\eeq
is 338 MeV.  The corresponding diquark mass is
\beq
M[sq] = m_s^b + m_q^b - \frac{3 a}{m_s^b m_q^b} =
536.3 + 363.7 - 3 (49.3)(363.7/536.3) = 800 {\rm~MeV}~,
\eeq
implying a reduced mass of 545 MeV.  According to Fig.~1 of Ref.\
\cite{Karliner:2015ema}, this would lead to the prediction $\Delta E_{PS}
\simeq 280$ MeV, nearly 60 MeV below the observed value.  So it is quite
possible that \SP splittings for baryons with one heavy quark and at least
one strange quark have been underestimated using the method of Ref.\
\cite{Karliner:2015ema}.

Some recent data from Belle \cite{Kato:2016hca,Yelton:2016fqw} on excited
$\Xi_c$ states may help identification of the spins and parities of the last
three states listed in Table \ref{tab:xic}.  The width of the state at
2970 MeV is measured to be about 30 MeV, while those for the states at
3055 and 3080 MeV are seen to be about 7 and less than 6.3 MeV, respectively.
That suggests $J^P = 1/2^-$ for the state at 2970 MeV and two $3/2^-$
assignments or one of $3/2^-$ and one of $5/2^-$ for the two higher-mass
states.

\section{Predictions for \boldmath
$\Omega_b = \bss$ states \unboldmath\label{sec:omb}}

The proposed identification of the five LHCb excited $\Omega_c$
states allows us to speculate upon the properties of a similar
system consisting of a $b$ quark and a spin-1 $(ss)$ diquark.
Here the large mass of the $b$ quark implies that the linear approximation
to the masses in Eqs.~(\ref{eqn:10}--\ref{eqn:52}) should be much better,
so we shall use it with the following inputs.
\begin{itemize}

\item[(i)] The hyperfine parameter $c$ is set to zero.

\item[(ii)] The parameter $a_1$ is kept as in the \css\ system, as it
expresses the coefficient of ${\bold L} \cdot {\bold S_{(ss)}}$:
$a_1[\bss] = a_1[\css] = 26.95$ MeV.

\item[(iii)] The parameter $a_2$ is rescaled by the ratio of heavy quark
masses:\\ $a_2[\bss] = (1708.8/5041.8)(25.74) = 8.72$ MeV, where we have
taken the charm and bottom quark masses from Ref.\ \cite{Karliner:2015ema}.

\item[(iv)] The parameter $b$ is taken to have a range of $\pm 20$ MeV around
zero, as in Ref.\ \cite{Karliner:2015ema}.

\item[(v)] The \SP splitting is taken as unknown, given that the reduced mass
of the \bss\ system, about 900 MeV, is outside the range for which we feel
comfortable making an estimate.  It should be {\em roughly} of the order of
300 MeV.

\end{itemize}
\newpage

\begin{figure}
\begin{center}
\includegraphics[width=0.95\textwidth]{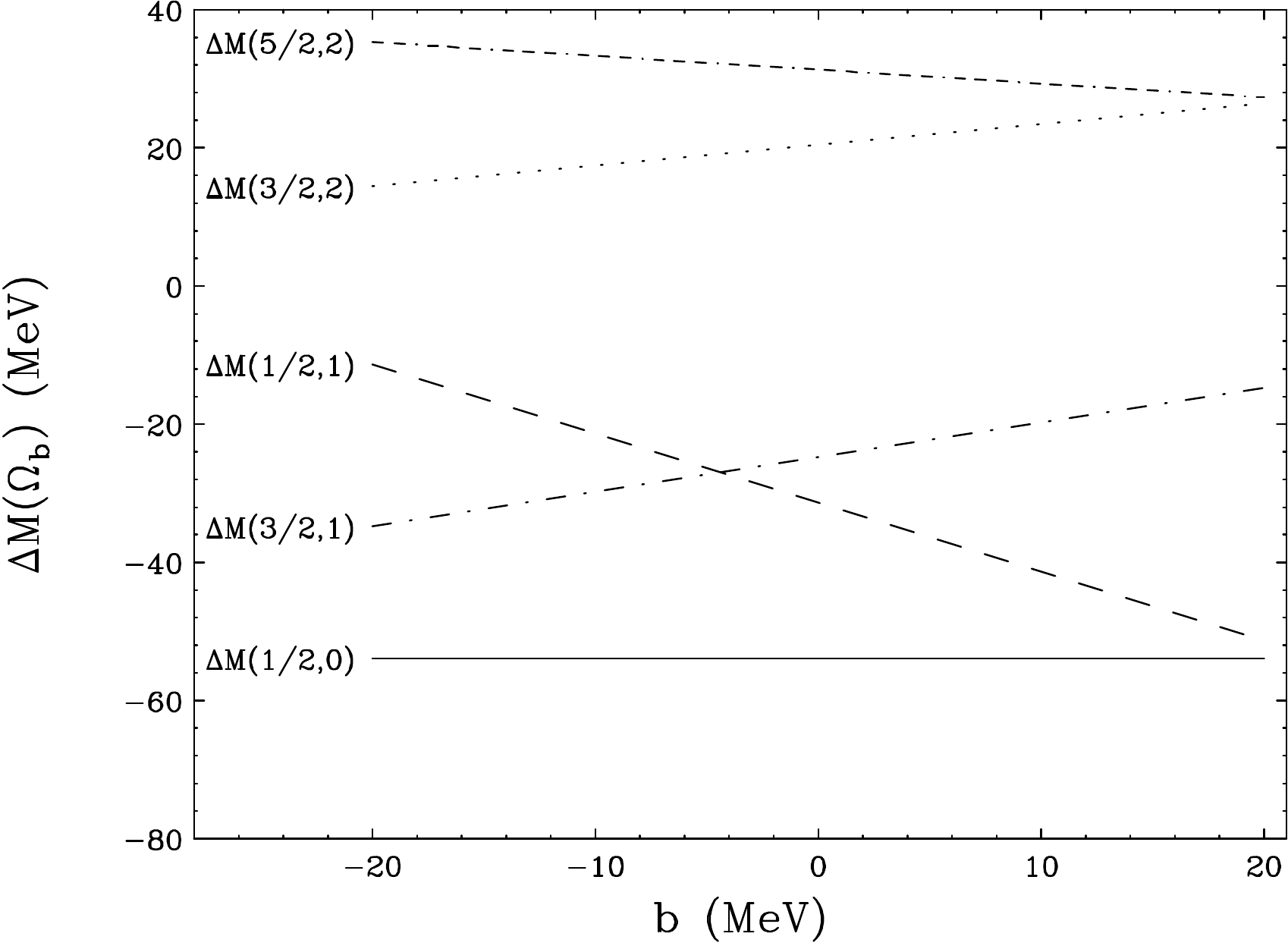}
\end{center}
\caption{Masses of $P$-wave $\Omega_b$ states \bss\ as functions of tensor
force parameter $b$ in scenario with all five peaks observed by LHCb
corresponding to $P$-wave excitations of the $(ss)$ diquark with respect to the
charmed quark.
\label{fig:omb}}
\end{figure}
These assumptions lead to the following mass shifts $\Delta M(J,j)$ in MeV
(see Fig.\ \ref{fig:omb}):
\bea
\Delta M(1/2,0) & = & -53.9~, 
\label{eqn:b10} \\
\Delta M(1/2,1) & = & -31.3 - b~, 
\label{eqn:b11} \\
\Delta M(3/2,1) & = & -24.8 + \frac12 b~, 
\label{eqn:b12} \\
\Delta M(3/2,2) & = &  20.4 + \frac{3}{10} b~, 
\label{eqn:b32} \\
\Delta M(5/2,2) & = &  31.3 - \frac15 b~.
\label{eqn:b52}
\eea
The order of the states is similar to that for the \css\
system, with only the shift $\Delta M(3/2,1)$ in indeterminate position
with regard to the shifts $\Delta M(1/2,0)$ and $\Delta M_(1/2,1)$.
As found in Ref.\ \cite{Karliner:2015ema} for the $P$-wave $\Sigma_b$ states,
for moderate $b$ there is a clear separation between the three lowest masses
with $j=0,1$ and the two highest with $j=2$.

The \,$\Omega_b^*$\, ($J^P=3/2^+$) \,partner of \,$\Omega_b(6046.4 \pm
1.9)$) \,should have a mass about
\break
\hbox{$(m_c/m_b)\Delta M(\Omega_c) \simeq (1/3)(71 {\rm~MeV}) \simeq
24$ MeV} above $\Omega_b$, so the spin-weighted average $S$-wave mass is about
6062 MeV.  The spin-weighted average of the five \bss\ states should then
be about 6362 MeV + $\Delta E_{PS}(\Omega_b) - 300$ MeV.

\section{\label{sec:alt}Alternative interpretations}

Predictions of $\sim 3000 \pm 40$ MeV for the negative-parity $\Omega_c$ states,
and an analogous range for the $\Omega_b$ states, have been made by several
authors \cite{Ebert:2011kk,Maltman:1980er,Ebert:2007nw,Roberts:2007ni,%
Garcilazo:2007eh,Migura:2006ep,Valcarce:2008dr,Yamaguchi:2014era,Bali:2015lka,%
Yoshida:2015tia,Shah:2016nxi,Wang:2017goq,Zhao:2017fov,
Agaev:2017jyt}.  (The last paper
treats only $2S$ levels, identifying the states at 3066, 3119 MeV as candidates
for $J^P = 1/2^+,3/2^+$, respectively.  In addition, simultaneously with or
after the first version of the present paper, there appeared works which also
identified the five observed $\Omega_c$ states as $1P$ excitations of the
$(ss)$ diquark with respect to the charmed quark \cite{Padmanath:2017,%
Wang:2017vnc,Wang:2017zjw,Chen:2017gnu,Aliev:2017led}, and interpretations
based on pentaquarks \cite{Yang:2017rpg,Huang:2017dwn,Kim:2017jpx}.)
The authors of Ref.\ \cite{Ebert:2011kk}
predict $M(1/2,1/2,3/2,3/2,5/2) = (3055,2966,3054,3029,3051)$ MeV for the
$P$-wave excited $\Omega_c$ states, and (6339,6330,6340,6331,6334) MeV for the
$P$-wave excited $\Omega_b$ states.  They, too, consider
only excitations in which the $(ss)$ diquark remains intact, with orbital
angular momentum 1 with respect to the heavy quark.  Most of the other excited
$\Omega_c$ predictions mentioned above are clustered somewhat below the
spin-weighted average based on our assignments in Table \ref{tab:omc}.

The possibility thus must be considered that not all of the states reported by
LHCb are $P$-wave excitations of the $(ss)$ diquark with respect to the charmed
quark \cite{Ebert:2011kk,Chen:2017sci,Wang:2017hej,Cheng:2017ove}).
Indeed, Ref.\ \cite{Ebert:2011kk} predicts candidates for the 2S
    \css\ states at 3088 MeV ($J^P = 1/2^+$) and 3123 MeV ($J^P = 3/2^+$),
not far from the two highest masses (3090 and 3119 MeV) reported by LHCb.
This leaves the states at 3000, 3050, and 3066 MeV to be identified as three
out of the five expected $P$-waves.  Where are the other two?

One possibility is that two of the observed peaks, though they
appear consistent with a single resonance, are actually composed of two, as
suggested by the near-degeneracies predicted in Ref.\ \cite{Ebert:2011kk}.
A spin-parity analysis of the LHCb data should resolve this question.

Another possibility is that one or both of the missing states are below
$\Xi_c^+ K^-$ threshold $(\simeq 2962$ MeV).  Such states would then be
expected to decay either by an electric dipole transition to $\Omega_c \gamma$
or via isospin violation/mixing to $\Omega_c \pi^0$ (in the manner of
$D_s(2317)$ decay).  The $\Omega_c \gamma$ spectrum has been studied by BaBar
\cite{Aubert:2006je} and Belle \cite{Solovieva:2008fw} in reporting the
existence of the $\Omega_c^{*0}$, a candidate for the $J^P = 3/2^+$ partner of
the $\Omega_c^0$.  The BaBar spectrum shows no peak above the $\Omega_c^{*0}$,
up to a mass of 3 GeV, while Belle only presents a spectrum up to an excitation
energy of 0.2 GeV, again showing no peak besides the $\Omega_c^{*0}$.  Still,
it might be interesting to examine the $\Omega_c \gamma$ and $\Omega_c \pi^0$
spectra in the forthcoming operation of Belle II.

\begin{figure}
\begin{center}
\includegraphics[width=0.95\textwidth]{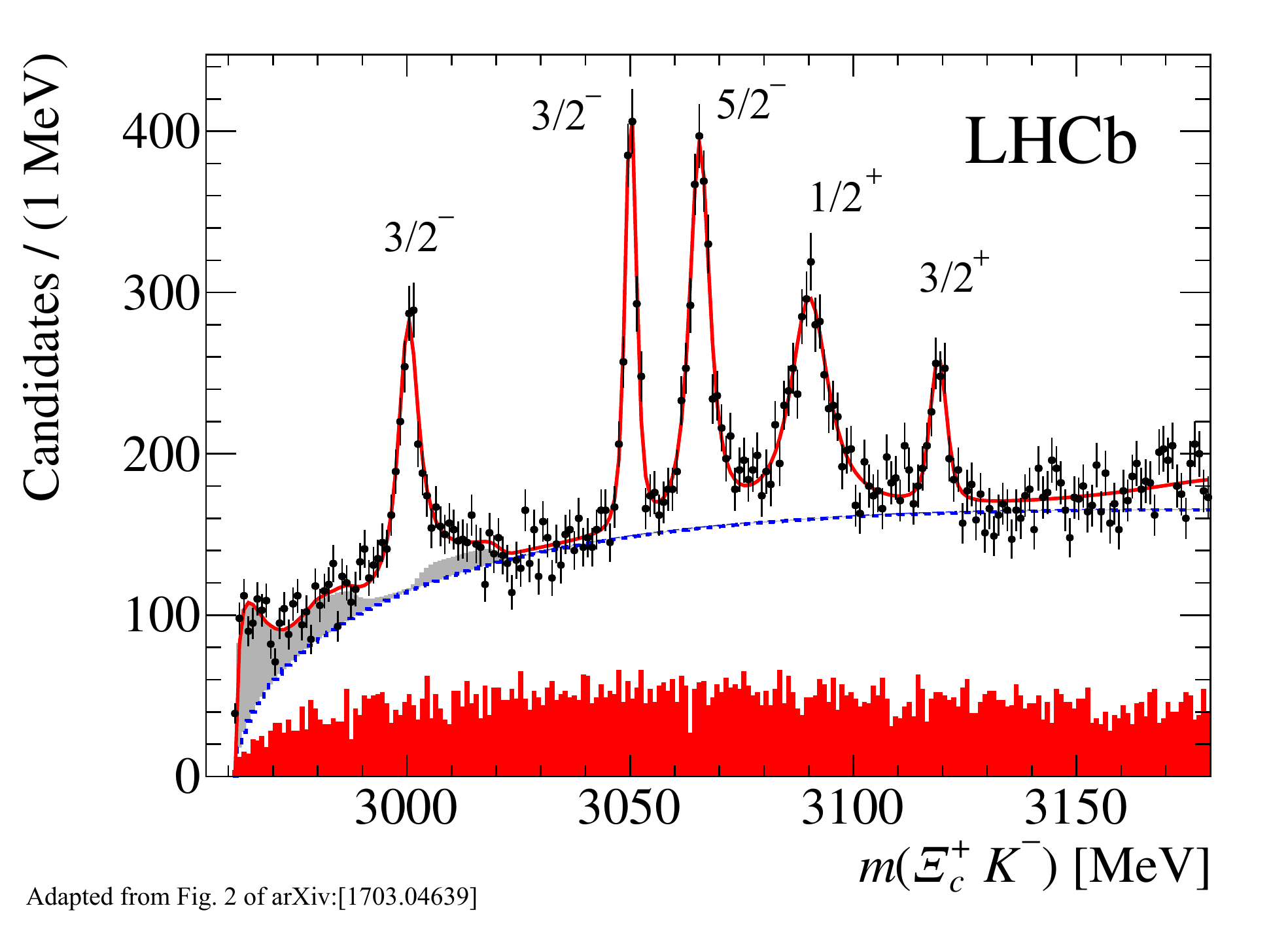}
\end{center}
\caption{\label{fig:xick1} Proposed assignment of spins and parities of excited
$\Omega_c = \css$ states observed by the LHCb Collaboration if lowest three
are $P$-wave excitations of the $(ss)$ diquark with respect to the charmed
quark, having $J^P = 3/2^-,3/2^-,5/2^-$, and upper two are $2S$ excitations
with $J^P = 1/2^+,3/2^+$.
Adapted from a zoom in on Fig.~2 of Ref.~\cite{LHCbOmegac}.}
\end{figure}

In a specific realization of this scenario, the states at 3000, 3050, and 3066
MeV are narrow because they decay via $D$ waves.  They then correspond to the
two states with $J^P = 3/2^-$ and the one with $J^P = 5/2^-$.  The two $J^P =
1/2^-$ states would be more elusive because either they are broader or they are
below $\Xi_c^+ K^-$ threshold.  To test this possibility, we choose parameters
motivated by the estimates in Sec.\ \ref{sec:sd}: $a_1 = 39.4$ MeV [item(iii)]
and $a_2=23.9$ MeV [item(ii)].
We vary $b$, $c$, and $\barM$ in a least-squares fit to the masses
$M(3/2,1)=3000.4$ MeV, $M(3/2,2) = 3065.6$ MeV, and $M(5/2,2) = 3119.1$ MeV.
We find $b = 27.85$ MeV, $c = -0.42$ MeV, and $\barM = 3020.03$ MeV, giving
rise to the predictions $M(1/2,0) = 2904.2$ MeV, $M(1/2,1) = 2978.0$ MeV.
(We thank Nilmani Mathur for informing us that this does not seem to be a
valid option in Ref.\ \cite{Padmanath:2017}.)
The corresponding alternative $J^P$ assignments of the LHCb peaks are shown in
Fig.\ \ref{fig:xick1}.  A fit with $M(3/2,2)$ and $M(5/2,2)$ interchanged
gives rise to an unphysical large negative value of $c$; other permutations
of the three negative-parity states do not result in a successful fit.

The favored solution's spin-averaged mass $\barM = 3020.03$ MeV is 278 MeV
above the $S$-wave spin-averaged value in Eq.~(\ref{Omega_c_spin_average}).
This is considerably closer to our estimate of about 240 MeV than the value of
about 338 MeV taking all five LHCb states as $P$-waves, 
Eqs.~\eqref{eq:Mbar} and \eqref{Omega_c_spin_average}.

The prediction of a state around 2904 MeV should be easy to confirm or refute
by studying the $\Omega_c \gamma$ and $\Omega_c \pi^0$ spectra.  As mentioned,
these were studied by Belle \cite{Solovieva:2008fw}, but only up to a mass of
about 2900 MeV, and by BaBar \cite{Aubert:2006je}, with no evidence for a
signal.
 
\begin{figure}
\begin{center}
\includegraphics[width=0.95\textwidth]{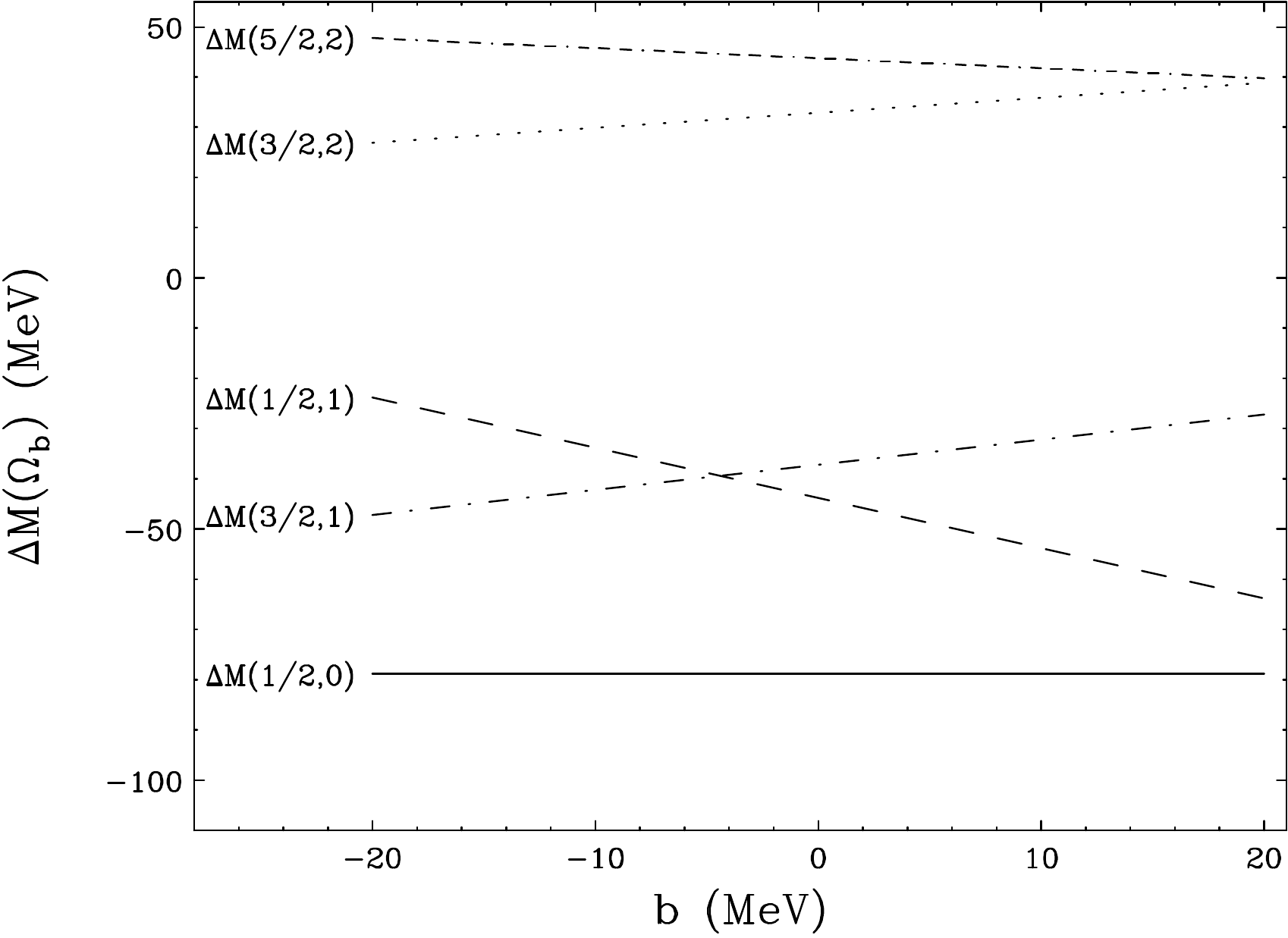}
\end{center}
\caption{Masses of $P$-wave $\Omega_b$ states \bss\ as functions
 of tensor force parameter $b$ in the scenario in which the LHCb peaks
have $J^P = 3/2^-,3/2^-,5/2^-,1/2^+,3/2^+$ in ascending order of mass.
\label{fig:omb1}}
\end{figure}

The predictions of the previous Section for $P$-wave $\Omega_b$ mass splittings
are altered in the present scenario, where we take $a_1 = 39.4$ MeV instead of
26.95 MeV [item (ii) of Sec.\ V].  The constants in Eqs.
     \eqref{eqn:b10}--\eqref{eqn:b52}
are replaced by $-78.8,-43.8,-37.2,32.9,43.8$ MeV, respectively, with the same
dependence on $b$.  The corresponding pattern of mass shifts qualitatively
resembles that of Fig.\ \ref{fig:omb}, with the $J^P = 5/2^-$ state and
one of the $J^P = 3/2^-$ states close to one another and significantly
heavier than the other three $P$-waves.  (See Fig.\ \ref{fig:omb1}.)

\section{Conclusions \label{sec:concl}}

The new excited $\Omega_c$ states observed by the LHCb Collaboration are a
spectroscopist's delight because of their high significance and narrow widths,
leading to well-defined and prominent signals.  We have interpreted these
five states in terms of the five states expected when a spin-1 $(ss)$ diquark
is excited with respect to the charm quark by one unit of orbital angular
momentum.  In our interpretation, the masses of the states are monotonically
increasing with their total spin.  This pattern remains to be confirmed.
If the two highest states instead are $2S$ with $J^P = 1/2^+$ and
$3/2^+$, the three lower states are likely on the basis of their narrow widths
to be two with $J^P = 3/2^-$ and one with $J^P = 5/2^-$. Then two predicted
$J^P = 1/2^-$ states remain to be identified, one around 2910 MeV decaying to
$\Omega_c \gamma$ and/or $\Omega_c \pi^0$ and the other around 2980 MeV
decaying to $\Xi_c^+ K^-$ in an $S$-wave.  We have also provided a template for
mass shifts in the corresponding \bss\ system.  It is not clear whether some
or all of the predicted $P$-wave states lie below $\Xi_b K$ threshold, in which
case they may be hard to identify, requiring identification in the $\Omega_b
\gamma$ or $\Omega_b \pi^0$ channel.  

\strut\vskip-1.3cm\strut
\section*{Acknowledgements}

We thank Simon Eidelman, Shmuel Nussinov, Antimo Palano, Marco Pappagallo,
and Tomasz Skwarnicki for helpful comments, Nilmani Mathur for alerting
us to Ref.\ \cite{Padmanath:2017}, and Richard Lebed for pointing out an
error in an earlier version of this manuscript.
The work of J.L.R. was supported by the U.S. Department of Energy, 
Division of High Energy Physics, Grant No.\ DE-FG02-13ER41958.

\section*{Appendix A}

An error in Eq.\ (A.1) of Ref.\ \cite{Karliner:2015ema} affects
the calculation of the tensor force.  The correct expression for $S_{12}/2$
[twice the contribution to Eq.\ (\ref{eqn:vsd})]~,
\beq
\frac{S_{12}}{2} \equiv
\langle 6 (\SSd \cdot {\bold r})
({\bold S_Q} \cdot {\bold r})/r^2 - 2 \SSd \cdot {\bold S_Q}\rangle~,
\eeq
is {\em not} equal to
\beq \label{eqn:Sdotr}
\langle 3( {\bold S} \cdot {\bold r})( {\bold S} \cdot {\bold r})/r^2
- {\bold S}^2 \rangle
\eeq
because a term in the latter expression quadratic in $\Sd$ does not vanish.
Instead, one has
\beq
\frac{S_{12}}{2} = \langle 3( {\bold S} \cdot {\bold r})( {\bold S} \cdot
{\bold r})/r^2 - {\bold S}^2 \rangle - {\cal C}~,~~{\rm where}
\eeq
\beq
{\cal C} = \langle 3( \SSd\cdot {\bold r})( \SSd
\cdot {\bold r})/r^2 - \SSd^2\rangle~.
\eeq
We now evaluate the correction term.  In Eqs. (A.4) and (A.5) of Ref.\
\cite{Karliner:2015ema},
we substitute $S \to \Sd$ and $J \to j$, with the result for $(j=0,1,2)$ that
${\cal C} = (-2,1,-1/5)$.  We want the matrix elements of ${\cal C}$ between
states $^{2S+1}L_J$, so we need the inverse of the Clebsch-Gordan relations
(A.15)--(A.18) of Ref.\ \cite{Karliner:2015ema}:
\bea
|^2P_{1/2}\rangle & = & \sqrt{1/3} |j=0 \rangle + \sqrt{2/3} |j=1 \rangle~,\\
|^4P_{1/2}\rangle & = & \sqrt{2/3} |j=0 \rangle - \sqrt{1/3} |j=1 \rangle~,\\
|^2P_{3/2}\rangle & = & \sqrt{1/6} |j=1 \rangle + \sqrt{5/6} |j=2 \rangle~,\\
|^4P_{3/2}\rangle & = & \sqrt{5/6} |j=1 \rangle - \sqrt{1/6} |j=2 \rangle~.
\eea

\bea
\bra{^2P_{1/2}}{\cal C}\ket{^2P_{1/2}}&=&\frac13(-2)+ \frac23(1)=0~,\\
\bra{^2P_{1/2}}{\cal C}\ket{^4P_{1/2}}&=&\sqrt2(-\frac23 -\frac13)=-\sqrt2~,\\
\bra{^4P_{1/2}}{\cal C}\ket{^2P_{1/2}}&=&-\sqrt2~,\\
\bra{^4P_{1/2}}{\cal C}\ket{^4P_{1/2}}&=&\frac23(-2)+ \frac13(1)=-1~,\\
\bra{^2P_{3/2}}{\cal C}\ket{^2P_{3/2}}&=&\frac16(1) + \frac56(-\frac15)=0~,\\
\bra{^2P_{3/2}}{\cal C}\ket{^4P_{3/2}}&=&\sqrt{5}(\frac16+\frac{1}{30})
 = \sqrt{5}/5~,\\
\bra{^4P_{3/2}}{\cal C}\ket{^2P_{3/2}}&=&\sqrt{5}/5~,\\
\bra{^4P_{3/2}}{\cal C}\ket{^4P_{3/2}}&=&\frac56(1)-\frac{1}{30}=\frac45~.
\eea

In addition a correction term
\beq
\bra{^4P_{5/2}}{\cal C}\ket{^4P_{5/2}} = - \frac15
\eeq
affects the contribution of the tensor force to ${\cal M}_{5/2}.$
The corrected mass operators are as shown in Sec.\ \ref{sec:sd}.

For completeness we describe here an alternative method of 
computing the tensor term. Denoting
$\bold n \equiv \bold r/r$, we have
\bea \label{Beqn:vsd}
\hat B &\equiv&
\langle
  - \SSd \cdot {\bold S_Q}
 + 3(\SSd \cdot {\bold n})({\bold S_Q} \cdot {\bold n})
\rangle
= 
3\langle
 n^i \Sdi n^j \SQj - \third\delta_{ij} \Sdi\SQj 
\rangle
\nonumber \\
\nonumber \\
&=& 3 \langle n^i n^j - \third \delta_{ij} \rangle \Sdi\SQj
\eea
Using the formula in Ref.~\cite{LL}
\beq
\langle n^i n^j - \third \delta_{ij} \rangle = 
a \left[ 
L_i L_j + L_j L_i - {\textstyle\frac{2}{3}}\delta_{ij} L(L+1)
\right],
\qquad
a = {-}1/[(2L-1)(2L+3)]
\eeq
for $L=1$ we have
\beq
\langle n^i n^j - \third \delta_{ij} \rangle = 
{-}{\textstyle\frac{1}{5}} 
\left(
L_i L_j + L_j L_i - {\textstyle\frac{4}{3}}\delta_{ij}
\right)
\eeq
so that
\bea
\hat B &=& {-} {\textstyle\frac{3}{5}}
\left(
L_i L_j + L_j L_i - {\textstyle\frac{4}{3}}\delta_{ij}
\right)
\Sdi\SQj
\nonumber\\
\nonumber\\
&=& 
{-} {\textstyle\frac{3}{5}}
\left(
L_i L_j \Sdi\SQj +
L_j L_i \Sdi\SQj 
- {\textstyle\frac{4}{3}} \SSd \cdot {\bold S_Q}
\right)
\nonumber\\
\nonumber\\
&=& 
{-} {\textstyle\frac{3}{5}}
\left[
(\bold L \cdot \SSd) (\bold L \cdot \bold S_Q)
+
(\bold L \cdot \bold S_Q) (\bold L \cdot \SSd) 
- {\textstyle\frac{4}{3}} \SSd \cdot {\bold S_Q}
\right]
\eea
where the last step is possible because 
$[\bold L, \SSd] = 
 [\bold L, \bold S_Q] = 
 [\SSd, \bold S_Q] = 0$.
Next, we want to compute matrix elements of $\hat B$ between states of
$J=1/2$, $J=3/2$ and $J=5/2$.
This can easily be done in terms of the known matrix elements of the 
three other operators,
$\hat A^1 \equiv \bold L \cdot \SSd$, 
$\hat A^2 \equiv \bold L \cdot \bold S_Q$ and 
$\hat C \equiv \SSd \cdot {\bold S_Q}$.
For example, the matrix elements of
$(\bold L \cdot \SSd) (\bold L \cdot \bold S_Q)$ can be computed
by inserting a complete set of states
between $\bold L \cdot \SSd$ and  $\bold L \cdot \bold S_Q$:
\beq
\bra\alpha
(\bold L \cdot \SSd) (\bold L \cdot \bold S_Q)
\ket\beta
=
\bra\alpha
(\bold L \cdot \SSd) \ket\gamma\bra\gamma(\bold L \cdot \bold S_Q)
\ket\beta
\eeq
Then 
\bea
\hat B_{J} &=&
{-} {\textstyle\frac{3}{5}}
\left(
\hat A^1_J \cdot \hat A^2_J +
\hat A^2_J \cdot \hat A^1_J
- {\textstyle\frac{4}{3}} \hat C_J
\right)
\eea
Explicitly,
\bea
\hat B_{1/2} &=&
{-} {\textstyle\frac{3}{5}}
\left({\textstyle
\left[ \begin{array}{c c} 
{-}\frac43  & {-}\frac{\sqrt{2}}{3} \\ 
\upstrut
{-}\frac{\sqrt{2}}{3} & {-}\frac53
\end{array} \right] 
\left[ \begin{array}{c c} 
\frac13 & \frac{\sqrt{2}}{3} \\ 
\upstrut
\frac{\sqrt{2}}{3} & {-}\frac56 
\end{array} \right]
+
\left[ \begin{array}{c c} 
\frac13 & \frac{\sqrt{2}}{3} \\ 
\upstrut
\frac{\sqrt{2}}{3} & {-}\frac56 
\end{array} \right]
\left[ \begin{array}{c c} 
{-}\frac43  & {-}\frac{\sqrt{2}}{3} \\ 
\upstrut
{-}\frac{\sqrt{2}}{3} & {-}\frac53
\end{array} \right] 
-\frac43
\left[ \begin{array}{c c} 
{-}1 & 0 \\
\upstrut
0 & \frac12
\end{array} \right]
}\right)
\nonumber\\
\nonumber\\
&=& 
\kern3em
\left[ \begin{array}{c c} 
0 & \frac1{\sqrt{2}} \\
\upstrut
\frac1{\sqrt{2}} & {-} 1
\end{array} \right]
\eea
\bea
\hat B_{3/2} &=&
{-} {\textstyle\frac{3}{5}}
\left({\textstyle
\left[ \begin{array}{c c} 
\frac23  & {-}\frac{\sqrt{5}}{3} \\ 
\upstrut
{-}\frac{\sqrt{5}}{3} & {-}\frac23
\end{array} \right] 
\left[ \begin{array}{c c} 
{-}\frac16 & \frac{\sqrt{5}}{3} \\ 
\upstrut
\frac{\sqrt{5}}{3} & {-}\frac13 
\end{array} \right]
+
\left[ \begin{array}{c c} 
{-}\frac16 & \frac{\sqrt{5}}{3} \\ 
\upstrut
\frac{\sqrt{5}}{3} & {-}\frac13 
\end{array} \right]
\left[ \begin{array}{c c} 
\frac23  & {-}\frac{\sqrt{5}}{3} \\ 
\upstrut
{-}\frac{\sqrt{5}}{3} & {-}\frac23
\end{array} \right] 
-\frac43
\left[ \begin{array}{c c} 
{-}1 & 0 \\
\upstrut
0 & \frac12
\end{array} \right]
}\right)
\nonumber\\
\nonumber\\
&=& 
\kern3em
\left[ \begin{array}{c c} 
0 & {-}\frac{\sqrt{5}}{10} \\
\upstrut
{-}\frac{\sqrt{5}}{10} & \frac45
\end{array} \right]
\eea
\beq
\hat B_{5/2} = {-}\frac35( 1\cdot\frac12 + \frac12\cdot 1
-\frac43\cdot\frac12)
={-}\frac15
\eeq

\section*{Appendix B}
The linearized approximation for the mass shift can be derived starting
from the exact expressions for $J=1/2$, $J=3/2$ and $J=5/2$:
\nl
\beq \label{Aeqn:m12}
\Delta {\cal M}_{1/2} = \left[ \begin{array}{c c} \frac13 a_2 - \frac43 a_1
&
\frac{\sqrt{2}}{3} (a_2-a_1) \\ \frac{\sqrt{2}}{3}(a_2-a_1) &
 - \frac53 a_1 - \frac56 a_2
\end{array} \right] +b \left[ \begin{array}{c c} 0 & \frac{1}{\sqrt{2}} \\
\frac{1}{\sqrt{2}}& -1 \end{array} \right] + c \left[ \begin{array}{c c} -1
&
 0 \\ 0 & \frac12  \end{array} \right]~,
\eeq
\nl
\beq \label{Aeqn:m32}
\Delta {\cal M}_{3/2} = \left[ \begin{array}{c c} \frac23 a_1 - \frac16 a_2
&
\frac{\sqrt{5}}{3}(a_2-a_1) \\ \frac{\sqrt{5}}{3}(a_2-a_1) &
 - \frac23 a_1 - \frac13 a_2
\end{array} \right] +b \left[ \begin{array}{c c} 0 & -\sqrt{5}/10 \\
 -\sqrt{5}/10 & \frac45 \end{array} \right] + c \left[ \begin{array}{c c}
-1 &
 0 \\ 0 & \frac12 \end{array} \right]~,
\eeq
\nl
\beq \label{Aeqn:m52}
\Delta {\cal M}_{5/2} = \ts a_1 + \frac12 a_2 - \frac15 b + \frac12 c~.
\eeq
\nl
States of definite $J$ and $j$ can be expressed as linear combinations of 
states with definite $J$ and $S$:
\nl
\nl
for $J=1/2$
\beq
\ket{J=1/2,j=0} = \sqrt{\ts\frac13}\ket{^2P_{1/2}}
                               + \sqrt{\ts\frac23}\ket{^4P_{1/2}}~,
\eeq
\beq
\ket{J=1/2,j=1} = \sqrt{\ts\frac23}\ket{^2P_{1/2}}
                               - \sqrt{\ts\frac13}\ket{^4P_{1/2}}~,
\eeq
for $J=3/2$
\beq
\ket{J=3/2,j=1} = \sqrt{\ts\frac16}\ket{^2P_{3/2}}
                               + \sqrt{\ts\frac56}\ket{^4P_{3/2}}~,
\eeq
\beq
\ket{J=3/2,j=2} = \sqrt{\ts\frac56}\ket{^2P_{3/2}}
                               - \sqrt{\ts\frac16}\ket{^4P_{3/2}}~.
\eeq
and for $J=5/2$
\beq
\ket{J=5/2,j=2} = \ket{^4P_{5/2}}~.
\kern7.7em\strut
\eeq
\nl
Then
\nl
\bea
\Delta M(J{=}\ts\frac12,j=0) &{=}& 
\bra{J{=}\ts\frac12,j{=}0} \Delta {\cal M}_{1/2} \ket{J{=}\ts\frac12,j{=}0} {=}
-2a_1~, \label{Aeqn:10}\\
\nonumber\\
\Delta M(J{=}\ts\frac12,j=1) &{=}& 
\bra{J{=}\ts\frac12,j{=}1} \Delta {\cal M}_{1/2} \ket{J{=}\ts\frac12,j{=}1} {=}
-\,a_1 -\ts\frac12 a_2 -b -\ts\frac12 c~,
\label{Aeqn:11} \\
\nonumber\\
\Delta M(J{=}\ts\frac32,j=1) &{=}& 
\bra{J{=}\ts\frac32,j{=}1} \Delta {\cal M}_{3/2} \ket{J{=}\ts\frac32,j{=}1} {=}
-\,a_1 +\ts\frac14 a_2 + \frac12 b + \frac14 c~,
\label{Aeqn:31} \\
\nonumber\\
\Delta M(J{=}\ts\frac32,j=2) &{=}& 
\bra{J{=}\ts\frac32,j{=}2} \Delta {\cal M}_{3/2} \ket{J{=}\ts\frac32,j{=}2} {=}
\ts \phantom{\strut-\strut}\,a_1 -\frac34 a_2 +\frac{3}{10} b - \frac34 c~,
\label{Aeqn:32} \\
\nonumber\\
\Delta M(J{=}\ts\frac52,j=2) &{=}& 
\bra{J{=}\ts\frac52,j{=}2} \Delta {\cal M}_{5/2} \ket{J{=}\ts\frac52,j{=}2} {=}
\ts \phantom{\strut-\strut}
\ts \,a_1 +\frac12 a_2 - \frac15 b + \frac12 c~.
\kern5em|
\label{Aeqn:52}
\eea

\end{document}